\documentclass[prb,twocolumn,amsmath,amssymb,english,superscriptaddress]{revtex4-1}

\usepackage[dvipdfm]{graphicx}
\usepackage{natbib}
\usepackage{subfigure}
\usepackage{tabularx}
\usepackage{epsfig}
\usepackage{longtable}
\usepackage{amsfonts}
\usepackage{rotating}

\usepackage{hyperref}
\usepackage{babel}

\usepackage{currfile}

\usepackage{color}

\newcommand{\bea}{\begin{eqnarray}}
\newcommand{\eea}{\end{eqnarray}}
\newcommand{\la}{\label}
\newcommand{\be}{\begin{equation}}
\newcommand{\ee}{\end{equation}}



\begin{document}

\title{FQHE interferometers in strong tunneling regime.
\\ The role of compactness of edge fields
 }

\author{Sriram Ganeshan, Alexander G. Abanov, Dmitri V. Averin}

\affiliation{Department of Physics and Astronomy, Stony Brook University,
Stony Brook, NY 11794-3800}

\date{\today}

\begin{abstract}

We consider multiple-point tunneling in the interferometers formed between edges of electron liquids with
in general different filling factors in the regime of the Fractional Quantum Hall effect (FQHE). We derive
an effective matrix Caldeira-Leggett model for the multiple tunneling contacts connecting the chiral single-mode FQHE edges. It is shown that the compactness of the Wen-Fr\"ohlich chiral boson fields describing the FQHE edge modes plays a crucial role in eliminating the spurious non-locality of
the electron transport properties of the FQHE interferometers arising in the regime of strong tunneling.


\end{abstract}

\maketitle


\section{Introduction.}
The defining feature of quantum Hall states is the existence of topologically protected massless edge states.
These states are believed to be effectively described by a theory of chiral bosons also known as the
one-dimensional chiral Luttinger Liquid ($\chi$LL) \cite{wenedge,1991-FrohlichKerler} (for review see \cite{2003-Chang}).
The tunneling experiments provide one of the natural ways to probe these edge states \cite{2003-Chang}.
The theory of edge-state tunneling was extensively developed over last two decades \cite{wenedge,FendleyLudwigSaleur,chamon,chamon0,chamon2,shytovprl,shytovprb,lopezfradkin,2003-Chang,averinjetp,averinprb1, averinantidot,feldman,averinprl,sukhorukov,chalker0,chalker1}.

The simplest model which is believed to universally describe a point contact between two single-mode FQHE edges consists
of two chiral bosons coupled by a tunneling cosine term. If both chiral edges are identical, the model can be mapped to
an integrable boundary sine-Gordon model and analyzed for arbitrary coupling strength between the edges, from the
limit of weak to strong tunneling \cite{FendleyLudwigSaleur}. In the limit of strong tunneling, the charge transfer
between chiral edges can be described by the instanton configurations of the model. The corresponding instanton expansion
can be understood as a weak coupling expansion of a dual model \cite{kanefisher,fradkin}.

The subject of this work is the system of chiral edges coupled via several point contacts. In the coherent regime,
the presence of several tunneling points allows for quantum interference between several paths of charge propagation.
In general, (with the exception of symmetric Mach-Zender interferometer \cite{averinmziprb,averinmzifcs1,averinmzifcs2})
the model with multiple tunneling contacts is not integrable and cannot be solved by Bethe Ansatz.

\section{Model.}
Let us start by defining the field theory model for FQHE interferometers. In this paper we focus on the case of a
two point-contact interferometer leaving generalizations to a multiple-point-contact case for the future. We consider
two chiral FQH edges corresponding to filling factors $\nu_{1,2}$. The Lagrangian for two FQH edges in imaginary time
formalism can be written in the bosonized form as
\bea
	L_{0}&=&\sum_{\sigma=1}^{2}\frac{\nu_{\sigma}}{4\pi}
	(\partial_{x}\phi_{\sigma})(i\gamma_\sigma\partial_{\tau}-v_\sigma \partial_{x})\phi_{\sigma} \, ,
 \label{freechiral}
\eea
where the fields $\phi_{1,2}$ are compact ($\phi_{\sigma}\equiv \phi_{\sigma}+2\pi$) chiral bosonic fields  describing two propagating chiral edge modes with velocities $v_{1,2}$ respectively.
The sign factors $\gamma_{1,2}=\pm1$ determine the direction of propagation of chiral fields (right/left), so that $\gamma_1\gamma_2=-1$ and $\gamma_1\gamma_2=1$  represent the cases of Fabry-Perot and Mach-Zehnder interferometers respectively. Each point contact tunneling can be modeled as a boundary Sine-Gordon term driven by the integrated vector potential $\Phi_{j}$ ($j=1,2$)
\begin{equation}
	L_{tunn}= -\sum_{j=1}^{2}\frac{U_j}{\delta}
	\cos\Big(\varphi(x_j,\tau)+\Phi_{j}(\tau)\Big)\,,
 \label{bsg}
\end{equation}
where
\begin{equation}
	\varphi(x,\tau)=\phi_1(x,\tau)-\phi_2(x,\tau)\,,
 \label{rotate}
\end{equation}
$U_{j}$ are tunneling amplitudes and $\delta$ represents the microscopic (ultraviolet) scale. The overall action of the system is
\be
	S=\int dx\,d\tau\,(L_0+L_{tunn}) \, ,
 \la{stot}
\ee
and the partition function of the interferometer is given by
\bea
	Z[\Phi_j] &=& \int D\phi_1 D\phi_2\, e^{-S[\phi_1,\phi_2,\Phi_j]} \,.
 \la{Zaj}
\eea
It is a functional of the e/m potentials $\Phi_{j}(t)$ which encodes electromagnetic responses of the interferometer ($\Phi_{j}$ is essentially the flux of magnetic field between wires to the left of the contact $j$). Namely, the variation of (\ref{Zaj}) with respect to $\Phi_{j}$ gives the tunneling current $I_{j}(\tau)$ through the contact $j$. A functional integration in (\ref{Zaj}) is taken over the compact bose fields $\phi_{j}\equiv \phi_{j}+2\pi$. We have chosen the normalization of the fields such that the electron operator on the edge $j$ is proportional to $e^{i\phi_{j}}$. In this work, we pay special attention to the compactness of the Bosonic fields $\phi_j$. Compactness is known to be important in the theory of non-chiral boson, where, in the path integral formulation, the compactness requirement is equivalent to allowing for instanton (vortex) configurations and might lead to the Kosterlitz-Thouless transition (see, e.g., \cite{PolyakovBook}). Quantum dynamics of mesoscopic Josephson junctions provides another example where the compactness properties of the phase operator $\varphi$ have important physical consequences \cite{azl1985}. In the present context of chiral fields, the proper treatment of compactness is crucial for obtaining correct results in the limit of strong tunneling $U_{j}\to \infty$. Indeed, in this limit, it is natural to assume that the values of the bose field at tunneling points $\varphi(x_{j})$ are pinned to the minima of the cosine in (\ref{bsg}) and the expansion in the number of instantons corresponding to the transitions between these minima will result in the expansion in $1/U_{j}$. It turns out however, that this procedure is plagued by unphysical non-localities \cite{averinjetp,averinprb1,averinprb2}. A work around solution to the problem was found by V. Ponomarenko and one of the authors \cite{averinjetp,averinprl} who introduced auxiliary ``zero modes'' designed to kill the nonlocal terms to produce physically meaningful result. The main goal of this work is to show that the correct treatment of the compactness of the bosonic field $\varphi$ makes the use of auxiliary modes unnecessary and produces physical results identical to Ref.~\cite{averinprl}.

\section{Green's function.}
The tunneling part of the model (\ref{bsg}) depends only on a difference $\varphi$ of bose fields (\ref{rotate}).
We can take advantage of this fact replacing $\phi_{1}\to \varphi+\phi_{2}$ and then integrating out the field $\phi_{2}$. This leaves us with the partition function (\ref{Zaj}) given by $Z[\Phi_j] = \int D\varphi \, \exp\left\{S_{0}+S_{tunn}\right\}$ where the tunneling part of the action is given by $S_{tunn}=\int d\tau\,L_{tunn}$ with (\ref{bsg}) while the free part of the action is given in Fourier representation by
\begin{equation}
	S_0[\varphi]=-\frac{1}{2}\int\frac{d^{2}k}{(2\pi)^{2}}\,\varphi^{*}
	\tilde{G}^{-1}\varphi \,,
 \label{eff_chiral}
\end{equation}
where
\be
	\tilde{G}(\omega,k)=\frac{4\pi}{\nu}\frac{\widetilde{v}
	-ia\omega/k}{(v_1k-i\gamma_1\omega)(v_2k-i\gamma_2\omega)} \,.
 \la{GtildeF}
\ee
Here we defined 
\begin{equation}
	\nu^{-1}=\frac{\nu_1^{-1}+\nu_2^{-1}}{2},\
	\widetilde{v}=\frac{\nu_1v_1+\nu_2v_2}{\nu_1+\nu_2},\
	a=\frac{\gamma_1\nu_1+\gamma_2\nu_2}{\nu_1+\nu_2}\,.
\la{parameters}
\end{equation}
We can think of $\nu^{-1}$, $\tilde{v}$ and $a$ as of average inverse filling factor, velocity and chirality of edges. In these notations,  the Mach-Zehnder interferometer corresponds  to $a=\pm 1$, the Fabry-Perot interferometer with $\nu_{1}=\nu_{2}$ corresponds to $a=0$. All other values of the parameter $a$ are between $-1$ and $1$ and correspond to an asymmetric Fabry-Perot interferometer with $\nu_{1} \neq \nu_{2}$.

In the limit of weak coupling, the generating function $Z[\Phi_{j}]$ can be written as a series expansion in $U_{j}$ using the correlation function (\ref{GtildeF}) rewritten as
\bea
	\tilde{G}(\omega,x) & =& \frac{2\pi i}{\omega} \int \frac{dk}{2\pi}\  e^{ikx}
	\bigg[\frac{2 a\gamma_1\gamma_2}{\nu k}
	-\sum_{\sigma=1}^{2}
	\frac{\gamma_{\sigma}/\nu_{\sigma}}{k-i\frac{\gamma_{\sigma}}{v_{\sigma}}\omega}\bigg]\,.
 \label{propagator}
\eea
We can split this propagator into local and non-local contributions $\tilde{G}=\tilde{G}_{nl}+\tilde{G}_{loc}$. The non-local contribution coming from the first term in Eq.~\ref{propagator} can be written as
\bea
	\tilde{G}_{nl}(\tau,x)&=& i\frac{\pi a\gamma_1\gamma_2}{\nu} sgn(\tau)\, sgn(x) \,.
\label{nonlocal}
\eea
Since $a \gamma_{1}\gamma_{2}/\nu = (\gamma_{1}\nu_{1}^{-1}+\gamma_{2}\nu_{2}^{-1})/2$ is strictly integer number, it contributes only as an overall phase to the correlation function of corresponding vertex operators $\langle e^{i\varphi(0,0)}e^{-i\varphi(\tau,x)}\rangle=e^{i\pi\frac{a}{\nu}}\exp\{-G_{loc}(\tau,x)\}$. The only non-trivial contribution comes from the local part of the correlation function given by the second term in (\ref{propagator})
\bea
	\tilde{G}_{loc}(\omega,x) = \sum_{\sigma=1}^{2}\frac{2\pi }{\nu_{\sigma} |\omega|}
               e^{-|\omega \frac{x}{v_{\sigma}}|} \theta(\omega \gamma_{\sigma}\frac{x}{v_{\sigma}})\,,
 \la{Gomegax}
\eea
At coinciding points we understand (\ref{Gomegax}) taking  $\theta(0)=1/2$.

Our goal is to find the partition function of the model (\ref{stot},\ref{bsg},\ref{eff_chiral}) in the limit of strong tunneling $U_{j}\to \infty$. First of all, following previous approaches \cite{caldeira,kanefisher} we are going to integrate our the degrees of freedom corresponding to the one-dimensional bulk and leave an effective action of Caldeira-Leggett type (CL) depending only on the values of
fields at the tunneling contacts.

\section{Effective Caldeira-Leggett model.}
The action of the tunneling model is quadratic except for the tunneling part (\ref{bsg}) localized at tunneling point. To integrate out the bulk field we impose the constraint
\begin{equation}
	 \varphi(x_j,\tau)=\varphi_j(\tau)\ \text{mod}\  2\pi
 \la{perconstr}
\end{equation}
by inserting
\be
	1=\int D\varphi_{j} \prod_j\delta_{P}(\varphi(x_j)-\varphi_j)
 \la{unres}
\ee
into the path integral. Here we introduced new ``contact'' fields $\phi_{j}(\tau)$ which are constrained to be the values of the $\varphi(x,\tau)$ field at contact  points $x=x_{j}$. The notation $\delta_{P}$ is chosen for the periodic version of the Dirac's delta function defined as
 \begin{equation}
	\delta_{P}(\varphi_j(\tau))=\sum_{\alpha_j(\tau)}e^{i\int d\tau \alpha_j(\tau)\varphi_j(\tau)}\,,
 \label{delta}
\end{equation}
where the sum is taken over the integer-valued fields $\alpha_{j}(\tau)$. \footnote{One should think of $\alpha_{j}(\tau)$ as of integer number defined for every slice of imaginary time.} The periodicity of the constraint (\ref{perconstr}) is the key point of our derivation. It is very important for the following and is a direct manifestation of the compactness of $\varphi$ which is ultimately related to the discreteness of electric charge.

%

Inserting (\ref{unres}) into the partition function and using (\ref{delta}) we can integrate out the field $\varphi$ and obtain
\bea
	Z[\Phi_{j}] &=& \int [D\varphi_{j}] \exp\left\{-S_{CL}[\phi_{j}]-S_{tunn}[\phi_{j},\Phi_{j}]\right\} \,
 \la{partPhi}\\
 	S_{tunn} &=& -\int d\tau\, \sum_{j}\frac{U_j}{\delta}
	\cos\Big(\varphi_{j}(\tau)+\Phi_{j}(\tau)\Big)
 \la{Stunn} \\
 	S_{CL} &=& -\log \int [D\alpha_{j}] e^{-\sum_{\omega}\alpha_i^{*}G_{ij}\alpha_j}
	e^{-i\sum_{\omega}\alpha_j^{*}\varphi_j} \,.
 \la{SCL}
\eea
Here the latter formula is written in frequency representation and we assumed the summation over tunneling points $i,j$. The CL matrix Green's function is defined in terms of (\ref{Gomegax})
as
\be
	G_{ij}(\omega) = \tilde{G}_{loc}(\omega,x_{i}-x_{j})\,.
 \la{Gij}
\ee
Notice that we dropped here the nonlocal part (\ref{nonlocal}) of the Green's function (\ref{propagator}). This is allowed as the $\alpha$ fields are integer-valued and the contribution to the first factor in (\ref{SCL}) is the overall phase. The integer valued fields $\alpha_j(\tau)$ imposes compactness condition on the point bosonic fields $\varphi_j(\tau)$ which becomes evident via Poisson summation formula\footnote{$\sum_{m=0}^{\infty}e^{\frac{-t}{2}m^2+ixm}=\frac{2\pi}{t}\sum_{n=0}^{\infty}e^{\frac{-1}{2t}(x-2\pi n)^2}$}. 

After dropping the oscillatory non-local part of the propagator we assume that the path integral (\ref{SCL}) is dominated by large fluctuations of $\alpha$ fields. We replace the functional summation over discrete $\alpha$ fields by Gaussian integration in (\ref{SCL}) and obtain
\bea
	S_{CL} &=&  \frac{1}{2}\sum_{\omega}\varphi_i^{*}(G^{-1})_{ij}\varphi_j \,.
 \la{SCL2}
\eea
Indeed, the dominant fluctuations of $\alpha$ fields can be estimated from  $\sum_{\omega}\delta\alpha_{i}^2(\omega)\approx\sum_{\omega}G_{ii}(\omega)\approx\sum_{\omega}|\omega|$, which is a huge number controlled by the ultraviolet cutoff of the problem. The partition function can be written as (\ref{partPhi}) with (\ref{Stunn},\ref{SCL2}), where
$(G^{-1})_{ij}$ is given by the matrix inversion of (\ref{Gij}).

\section{Current in weak tunneling limit.}
To calculate the tunneling current in the limit of weak tunneling we need the correlators of vertex operators given by \footnote{Note here that one could use instead of (\ref{Gij}) the Green's function including the nonlocal term (\ref{nonlocal}). This would not change the correlation functions for vertex operators.}
\bea
	\langle e^{i\varphi_j(\tau)}e^{-i\varphi_k(0)}\rangle
	&=& e^{2\sum_{\omega}e^{i\omega \tau}G_{jk}}
 \nonumber \\
        &=& K(\xi^{1}_{jk},i\tau)K(\xi^{2}_{jk},i\tau) \,.
 \label{correl}
\eea
where the parameters related to different propagation times are defined as
\be
	\xi_{jk}^{\sigma}=\gamma_{\sigma}\frac{x_j-x_k}{v_{\sigma}},\;\; \xi^{\sigma}=|\xi_{jk}^{\sigma}|,\;\;
    \xi^{tot}= \sum_{\sigma}\xi^{\sigma} \,.
\ee
Here we defined
\begin{equation}
	K(\xi^{\sigma}_{jk},i\tau)
	=
                  \Big[\frac{ \pi T}{v_{\sigma} \sinh{\pi T (\xi^{\sigma}_{jk}+i\tau)}}\Big]^{\frac{1}{\nu_{\sigma}}} \, ,
 \la{K}
\end{equation}
which in the limit of zero temperature $T=0$ gives a well known result $(v_{\sigma}(\xi^{\sigma}_{jk}+i\tau))^{-1/\nu_{\sigma}}$.

Using the correlation function (\ref{correl},\ref{K}) we obtain the tunneling current through the $j$-th point contact in the lowest order of perturbation theory as \cite{kanefisher}
\bea
	I_j(t)&=&\sum_{i}\frac{U_{i}U_{j}}{2\delta^2}\int_{-\infty}^{t}dt_1\sin(\Phi_i(t_1)-\Phi_j(t))	
 \nonumber\\
	&\times& Im\left( \prod^{2}_{\sigma=1}K(\xi^{\sigma}_{ij},t_1-t-i0)\right) \,.
 \la{pccurrentweak}
\eea
In fact, it is easy to write down the expression for the current at any point on the interferometer as a function of real time
\bea
	I^{\sigma}(x,t)&=&I_{cs}(E_{\sigma})+\sum_j\theta(\gamma_{\sigma}(x-x_j))I_j(t) \,,
 \la{edgecurrent}
\eea
where $I_{cs}=\frac{v_{\sigma}\nu_{\sigma}}{2\pi}\int_{-\infty}^{0} dt'E^{\sigma}(x+\gamma_{\sigma}v_{\sigma} t',t)$ is the current due to the electric field $E_{\sigma}$ along the edge due to the contribution of the bulk Hall current.

\section{Role of compactness in strong tunneling limit.}
In the limit of strong tunneling $U_{j}\to \infty$ the path integral in (\ref{partPhi}) will be dominated by fields pinned to the minima of tunneling cosine term (\ref{Stunn}) so that the optimal field configurations are given by
\be
	\varphi_j(\tau)= 2\pi n_j-\Phi_j(\tau) \,,
 \label{eq:ground}
\ee
where $n_{j}$ are arbitrary integer numbers. The calculation of the generating functional (\ref{partPhi}) in this limit amounts to substitution of the optimal field configurations (\ref{eq:ground}) into (\ref{partPhi},\ref{Stunn},\ref{SCL}). It is very important that although the nonlocal contribution to Green's function (\ref{nonlocal}) would be of no consequence in calculating (\ref{SCL}) and in (\ref{correl}) it is vital to use only the local expression (\ref{Gij},\ref{Gomegax}) if one calculates the value of Caldeira-Leggett action using (\ref{SCL2}) with the inverse of the local Green's function (\ref{Gomegax},\ref{Gij}) given by
\begin{equation}
	(G^{-1})_{ij}
	=  \frac{\nu|\omega|}{2\pi D(\omega)}
 \left\{\begin{array}{ll}
		1 & (i=j)
		 \\ \\
                  -\sum_{\sigma=1}^{2}\frac{\nu}{\nu_{\sigma}}e^{-|\omega \xi^{\sigma}|}\theta(\omega \xi^{\sigma}_{ij})
                  & (i \ne j) \\
                       \end{array} \right. \,,
 \la{G}
\end{equation}
where we defined the resonant denominator as
\be
	D(\omega)=(1-(1-a^2)e^{-\sum_{\sigma}|\omega \xi^{\sigma}|}) \,
\ee
with $a$ given by (\ref{parameters}).

\section{Current in the limit of infinitely strong tunneling.}
Let us substitute Eq.~(\ref{eq:ground}) in to Eq.~(\ref{SCL2}) and neglect fluctuations around the optimal configurations (\ref{eq:ground}). The terms coming from $n_j$ do not contribute to the tunneling current and hence can be dropped. In this regime the action and the partition function can be written as
\bea
	S_0&=&\frac{1}{2}\sum_{\omega,ij}\Phi^*_i(G^{-1})_{ij}\Phi_j \,.
\eea
By taking variation with respect to $\Phi^{*}_k$ and analytically continuing to real time  we obtain current through point contact
\be
	I^{\infty}_k(\omega)=\sum_j(G^{-1})_{kj}|_{ret}\Phi_j \,,
 \la{currentinf}
\ee
 where we introduced the retarded real time Green's function $G_{ret}$ \footnote{To obtain the retarded Green's function $(G^{-1})_{jk}|_{ret}$ we should analytically continue $i\omega\to\omega+i0$ which amounts to $|\omega|\to \omega$, $|\omega \xi^{\sigma}|\to -i\omega \xi^{\sigma}$ and $\theta(\omega\xi^{\sigma}_{ij})\to \theta(\xi^{\sigma}_{ij})$ in \ref{G}}. Using the expansion in Eq.~(\ref{eq:binomial}), we obtain from (\ref{currentinf}) the current through the point contacts as
\bea
	I^{\infty}_{1,2}(t)
	&=&\frac{\nu}{2\pi}\sum^{\infty}_{m=0}(1-a^2)^m \big[ \dot{\Phi}_{1,2}(t-m\xi^{tot})
 \nonumber\\
 	&-&\sum_{\sigma=1}^{2}\frac{\nu\theta(\mp\gamma_{\sigma})}{\nu_{\sigma}}\dot{\Phi}_{2,1} (t-\xi^{\sigma}-m\xi^{tot})
	\big] \,.
 \la{Iinf}
\eea
Consider the example when the constant voltage $V$ between wires is turned on at time $t=0$ ($\Phi_j(t)=Vt\theta(t)$). The expression (\ref{Iinf}) allows for a very straightforward interpretation in the picture of strong tunneling suggested in \cite{chamon}.  
\bea
	I^{\infty}_{1,2}(V,t)
	&=&\frac{V \nu}{2\pi}\sum^{\infty}_{m=0}(1-a^2)^m \big[ \theta(t-m\xi^{tot})
 \nonumber\\
 	&-&\sum_{\sigma=1}^{2}\frac{\nu\theta(\mp\gamma_{\sigma})}{\nu_{\sigma}}\theta (t-\xi^{\sigma}-m\xi^{tot})
	\big] \,.
 \la{Iinfv}
\eea
The summation over $m$ in (\ref{Iinf}) is due to the current traveling around the closed edge formed between contacts ($\xi^{tot}$ is then the total time required to travel around that loop). For the case of MZI ($a=1$) with constant voltage between the wires, only the term with $m=0$ contributes to (\ref{Iinf}) and at large times $t>\mbox{max}(\xi^{1},\xi^{2})$ we have $I^{\infty}_{1,2}=\pm\frac{\nu V}{2\pi}$. For the case of symmetric FPI ($a=0$), there is a feedback of current carried by one of the opposite chirality edges and the current switches between the value $\frac{\nu V}{2\pi}$ and $0$ with the overall period $\xi^{tot}$. We emphasize here that this physical picture of tunneling at $U=\infty$ limit is a direct consequence of correct treatment of compactness of chiral bosons.

\section{Instanton expansion and dual model.} In the limit of strong but finite tunneling the instanton corrections to the saddle point considered in the infinite tunneling limit become important. Namely, the bosonic fields will once in a while change their values to $2\pi(n_j\pm1)$
\be
	\varphi_j=2\pi n_j-\Phi_j(\tau)+2\pi\sum_l e^{l}_{j}\theta(\tau-\tau^{l}_{j}) \,.
 \label{eq:instanton}
\ee
Here $\tau^{l}_{j}$ is the time of switching of the value of $\varphi_{j}$ and
 $e^{l}_{j}=\pm$ is the sign of the switching corresponding to instanton/anti-instanton tunneling events. The summation over instantons and anti-instantons can be rewritten as a perturbative expansion in quasiparticle tunneling amplitudes $W_{j}$ \footnote{$W_{j}\sim 1/U_{j}$ is the action cost of an instanton.} of the following dual tunneling model \cite{averinprl,averinprb2,kanefisher}
\bea
	S [\Theta_1,\Theta_2]
	&=& \frac{1}{2}\sum_{\omega,ij}\Theta_i^{*}M^{-1}_{ij}\Theta_j
	+\frac{1}{2}\sum_{\omega,ij}\Phi^{d*}_iM^{-1}_{ij}\Phi^d_j
 \nonumber\\
 	&-&\sum_j\frac{W_j}{\delta}\int d\tau \cos\left(\Theta_j(\tau)+\Phi^d_j(\tau)\right) \,,
 \nonumber \\
 	M_{ij}&=& \left(\frac{4\pi}{\omega}\right)^2(G^{-1})_{ij} \,,
 \\
	 \Phi^d_j(\omega)&=&\frac{\omega}{2\pi}M_{ji}(\omega)
	 (\Phi_i(\omega)-2\pi n_i\delta(\omega)) \,.
     \label{dualgauge}
\eea
The partition function for the dual model reproduces the strong tunneling instanton expansion and in the lowest order in $W_{j}$ we have
\bea
	\frac{Z}{Z_0}
	&=&e^{-\frac{1}{2}\sum_{\omega,ij}\Phi^{d*}_iM^{-1}_{ij}\Phi^d_j}
	\Big[1+\frac{1}{4}\sum_{n_1,n_2}\sum_{ij}\frac{ W_i W_j}{\delta^2}
 \nonumber\\
           &\times & \int d\tau_1 d\tau_2\cos(\Phi^d_i(\tau_1)-\Phi^d_j(\tau_2))\langle
           e^{i\Theta_i(\tau_1)}e^{-i\Theta_j(\tau_2)}\rangle\Big]
 \nonumber
\eea
Two point quasiparticle correlators can be computed as,
\bea
	\langle e^{i\Theta_i(\tau)}e^{-i\Theta_j(0)}\rangle
	&=& e^{2\sum_{\omega}e^{i\omega \tau}M_{ij}}
 \label{correldual} \\
	&=&\prod^{\infty}_{m=0} K^{qp}(\chi^{1}_{ij}(m),i\tau)K^{qp}(\chi_{ij}^{2}(m),i\tau)\,,
 \nonumber
\eea
where we defined
\bea
	K^{qp}(\chi^{\sigma}_{ij}(m),i\tau)
	&=& \begin{array}{c}
                  \Big[\frac{ \pi T}{v_{\sigma}
                  \sinh{\pi T (\chi^{\sigma}_{ij}(m)+i\tau)}}\Big]^{\frac{\nu^2(1-a^2)^m}{\nu_{\sigma}}}
                 \end{array} \,,
 \nonumber \\
	\chi_{ij}^{\sigma}(m)
	&=& \left(\xi^{\sigma}(1-\delta_{ij})
	+m\xi^{tot}\right)~\text{sgn}(\xi^{\sigma}_{ij})
 \label{qpchiral}
\eea
and the product over $m$ in (\ref{correldual}) appears from the expansion
\be
	\frac{1}{D(\omega)}
	=\sum^{\infty}_{m=0}(1-a^2)^m~e^{-m\sum_{\sigma}|\omega \xi^{\sigma}|} \,.
\la{eq:binomial}
\ee
The above formulas are valid for both FP and MZ interferometers with different geometries encoded by the parameter $a$. The current at any point on the edge is given again by Eq.~(\ref{edgecurrent}) with the tunneling current at a point contact
\bea
	I_j(t)-I_j^{\infty}
	&=&\sum_{n_1,n_2}\sum_{k,i}\frac{W_{k}W_{i}}{2\delta^2}
	\int_{-\infty}^{t}dt_2\,\int_{-\infty}^{\infty}dt_1\,
 \la{pccurrentqp} \\
 	&& \sin\left(\Phi^d_k(t_1)-\Phi^d_i(t_2)\right)
	J_{kj}(t_1-t)
 \nonumber\\
	&\times& Im\left(\prod^{\infty}_{m=0}\prod^{2}_{\sigma=1}
	K^{qp}(\chi^{\sigma}_{ki}(m),t_1-t_2-i0)\right) \,,
 \nonumber
\eea
where $J_{kj}=\omega M_{kj}/2\pi$ is obtained from the variation $\delta \Phi^d_k/\delta \Phi_j$ calculated from (\ref{dualgauge}). We can calculate the explicit expression for the time dependent Jacobian for the general vector potentials
\be
	\Phi_i^{d}(t)=\int_{-\infty}^{\infty}dt_{1}\sum_{j}\text{J}_{ij}(t-t_{1})\Phi_j(t_{1})\,.
\ee
We can write the Jacobian matrix as,
\bea
	J_{i=j}(t) &=& 4\nu\sum_{m=0}^{\infty}(1-a^2)^m\delta(t-m\xi^{tot}) \,,
 \\
	J_{i \ne j}(t) &=& -4\nu\sum_{m=0}^{\infty}\sum_{\sigma=1}^{2}
	\frac{\nu(1-a^2)^m}{\nu_{\sigma}}\delta(t-\xi^{\sigma}-m\xi^{tot})\theta(\xi_{ij}^{\sigma}) \,.
 \nonumber 
\eea
The eq.~(\ref{pccurrentqp}) describes the correction to the current at infinite coupling $I_{j}^{\infty}$ due to quasiparticle tunneling processes (instantons) \cite{kanefisher,fradkin}.
We note here that the interference current (\ref{pccurrentqp}) has a $2\pi$ (electron) periodicity with respect to the flux $\Phi=\Phi_{2}-\Phi_{1}$ between tunneling contacts. This physical periodicity can be directly traced as coming from the compactness of the edge fields in (\ref{Zaj}).


\section{Conclusion.} We showed that compactness of the chiral bosonic fields in the tunneling model of FQHE interferometers plays a
crucial role in the limit of strong tunneling. Taking compactness of fields into account, we derived the effective Caldeira-Leggett model (\ref{partPhi},\ref{Stunn},\ref{SCL}) as an effective description of the two point interferometers.  We showed that this model reproduces the results for tunneling currents known in literature in both weak and strong tunneling limits and gave the expressions for those currents in a unified form with the chirality parameter $a$ from (\ref{parameters}) encoding the geometry of the interferometer (Fabry-Perot and Mach-Zehnder with arbitrary filling factors). Generalization of the proposed formalism to the multi-point interferometers should be straightforward.

\paragraph*{Acknowledgment.---}
The work of A.~G.~A.\ was supported by the NSF under Grant No.\ DMR-1206790. S.~G.\  was supported by DOE award number DE-FG02-09ER16052.






%


%


\end{document}